\documentclass[aps,prb,twocolumn,groupedaddress,showpacs,showkeys]{revtex4}

\usepackage{graphicx,amsmath,amssymb}
\nonstopmode
\newcommand{\ppp}{($+$)\,}
\newcommand{\pmp}{($-$)\,}
\def\<{\langle}
\def\>{\rangle}
\newcommand{\etal}{{\it et al}}
\newcommand{\uf}{^{4+}}
\newcommand{\ut}{^{3+}}
\newcommand{\ie}{{\it i. e.}}

\begin{document}
\preprint{cond-mat}
\title{
Studies of transverse and longitudinal relaxations of $^{55}$Mn
in molecular cluster magnet Mn$_{12}$Ac}

\author{Takao Goto}
\email{goto@nmr.jinkan.kyoto-u.ac.jp}
\author{Takeshi Koshiba}
\altaffiliation[Present address: ]{Toshiba Corporation, Semiconductor
Company, 8 Shinsugimoto-cho, Isogo-ku, Yokohama, 235-8522, Japan}
\affiliation{Graduate School of Human and Environmental Studies,
Kyoto University, Kyoto 606-8501, Japan}
\author{Takeji Kubo}
\affiliation{Physics Department, Faculty of Education, Nara
University of Education,Nara 630-8301, Japan}
\author{Kunio Awaga}
\affiliation{Department of chemistry, Graduate School of Science,
Nagoya University, Nagoya, 464-0814,
Japan}
\date{\today}

\begin{abstract}
The nuclear magnetic relaxation times $T_2$ and $T_1$ of $^{55}$Mn
in the molecular cluster magnet Mn$_{12}$ Ac have been measured,
using the spin-echo method for oriented powder sample,
at low temperatures below 2.5\,K down to 200\,mK
in the fields up to 9\,T applied  along the $c$-axis.
Above about
1.5\,K both of relaxation rates $T_{2}^{-1}$
and $T_{1}^{-1}$ exhibit remarkable
decreases with decreasing temperature in zero field
with the relative relation like $T_2^{-1}/T_1^{-1}\approx200$.
At the lower temperatures,
$T_{2}^{-1}$  tends to become
constant with the value of about
$10^2$\,s$^{-1}$, while $T_{1}^{-1}$ still exhibits
an appreciable
decrease down to around 0.5\,K.
The analysis for the experimental results was made on basis of the concept that
the fluctuating local field responsible for the nuclear magnetic
relaxation is caused by thermal fluctuations of the Zeeman levels of
the cluster spin of $S=10$ due to  the spin-phonon interactions.
Then the problem was simplified by considering
only
the thermal excitation from the ground state to the first
excited state, that is,   
a step-wise 
fluctuation with respective
average life times $\tau_0$ and $\tau_1$. 
By applying nonlinear theory 
for such a
fluctuating local field, a general expression for $T_{2}$ was obtained.
It turned out that the experimental results for  $T_{2}^{-1}$ are explained 
in terms of the equation of $T_2^{-1}=\tau_0^{-1}$, 
which corresponds to the strong collision regime 
under the  condition $\tau_0\gg\tau_1$.
On the other hand, the results for $T_1$ have
been well understood, on the basis of the standard
perturbation method, by the equation for the the high-frequency limit 
like $T_1^{-1}\sim 1/\tau_0 \omega_N^2$, where 
$\omega_N$ is the $^{55}$Mn Larmor frequency.  
The experimental results for the field dependence of the $T_2^{-1}$ and $T_1^{-1}$ 
were also interpreted reasonably in terms of the above theoretical 
treatment.
The quantitative comparison between the experimental results 
and the theoretical equations
was made using hyperfine interaction tensors for each of 
three manganese ions determined from the 
analysis for the NMR spectra in zero field.  
\end{abstract}

\pacs{75.50.Xx, 76.60.-k, 75.45.+1}
\keywords{Molecular magnet Mn$_{12}$Ac,
$^{55}$Mn NMR, Relaxation times $T_1$ and $T_2$}
\maketitle

\section{Introduction\label{sec:Intro}}
Recently there has been a great interest in nano-scale molecular
cluster magnets
in view of the  microscopic quantum nature appearing in the
macroscopic properties of the system. The  synthesis of the molecular
cluster magnets have solved the most serious problem of the particle size,
because the cluster size of the cluster magnet is exactly same and known.
As a typical candidate compound, the molecular cluster magnet
Mn$_{12}$O$_{12}$(CH$_3$COO)$_{6}$(H$_2$O)$_4$
(abbreviated as  Mn$_{12}$Ac),
which was synthesized  and whose crystal structure was studied using
X-ray diffraction by Lis,~\cite{L}
has been studied so far most extensively.
Magnetic properties of Mn$_{12}$Ac 
have been explained satisfactorily by treating the strongly-coupled cluster
spins as a single quantum spin of
$S$=10. 
The prominent features have been demonstrated in 
very long relaxation time of the magnetization at low temperatures,~\cite{SGCN,VBSP}
and 
the step-wise recovery of the magnetization in the external field 
at every interval of about 0.45\,T, 
which is associated with
the quantum tunneling which occurs with coincidence in
the Zeeman levels of oppositely directed magnetization of the cluster
spins.~\cite{TLBGSB,FSTZ}
Such a tunneling  phenomenon was
interpreted as thermally-assisted and/or field-tuned processes
associated with the spin-phonon interaction.~\cite{PRHV}

Another molecular compound [(tacn)$_6$Fe$_8$O$_2$(OH)$_{12}$]$^{8+}$
(abbreviated as Fe$_8$) with a large cluster-spin of  $S$=10 has been the
subject of investigations  on
the same viewpoint.~\cite{SOPSG,WSG}
On contrary to
Mn$_{12}$Ac and Fe$_8$ in which the tunneling is due to ``spin-bath'',
a  molecular cluster magnet
K$_8$[V$_{15}^{{\rm IV}}$As$_6$O$_{42}$(H$_2$O)]$\cdot$8H$_{2}$O (so called
V$_{15}$) has been studied as a system with a lowest cluster spin of $S$=1/2,
which shows a hysteresis curve  associated with ``phonon-bath'',
although there is no energy barrier against the spin reversals.~\cite{CWMBB}

A great deal of experimental and
theoretical works related to Mn$_{12}$Ac
has been reviewed in some books and review papers.~\cite{GB,CT,HPV,TPS,TB}

In order to understand the magnetic
properties of these  molecular cluster magnets more thoroughly, it is
worthwhile to examine the dynamical behavior as well as statistical
nature of each magnetic ion which constitutes the cluster spin.
One of the most promising experimental procedures for this
purpose will be to use an NMR with respect to
the relevant magnetic ions.
In view of this, we have been studying $^{55}$Mn NMR in
Mn$_{12}$Ac.
As reported in our preliminary paper,
we  first succeeded, using powder sample,
in observing all of $^{55}$Mn NMR signals belonging to Mn$\uf$
and two in-equivalent Mn$\ut$ ions in zero field,
and measured the temperature dependence of
the transverse relaxation time $T_2$
at liq. Helium temperatures.~\cite{GKKFOATA}
In the recent paper, we have determined the hyperfine interaction
tensors of $^{55}$Mn nuclei in Mn$_{12}$Ac by analyzing the more detailed
$^{55}$Mn NMR
spectra on the basis of the ground-state
spin-configuration.~\cite{KGKTA}

In the present work, we have measured, using oriented powder sample,
the transverse relaxation time $T_2$  and the spin-lattice relaxation
time $T_1$ of $^{55}$Mn in Mn$_{12}$Ac. The measurements have been
done in the wide temperature range from 2.5\,K down to 200\,mK in zero field and at
liq. Helium temperatures with an applied field up to 9T.  As for the
nuclear-spin lattice relaxation in Mn$_{12}$Ac,  the proton spin- and
muon spin-lattice relaxation times have been measured  by Lascialfari
\etal. at the temperature range of 4.2-400\,K and at the field  range
of 0-9.4\,T,~\cite{LGBSJC}  and subsequently  at the lower
temperatures down to around 2\,K below 1.6\,T.~\cite{LJBCG}
In Ref.~\onlinecite{LJBCG} the field
dependence of the proton relaxation rate $T_{1}^{-1}$ and the temperature
dependence of the muon relaxation rate 
have been
analyzed on the basis of the standard perturbation method, that is,
weak collision model.
Then the  fluctuating local
fields at the proton or the muon site  have been taken to  originate
from the random change between the adjacent  Zeeman-energy levels of
the ground-state configuration of the total spin of $S$=10, which is
caused by the spin-phonon interactions.
Quite recently Furukawa \etal. have reported on the experimental results
for the temperature
and field dependence of $T_{1}^{-1}$ of $^{55}$Mn in Mn$_{12}$Ac above
1.2\,K and below 1\,T.~\cite{FWKBG}, and the analysis has been made
on the basis of essentially the same treatment given
in Ref.~\onlinecite{LJBCG}.

The analysis for our experimental results are made 
on the basis of the concept as presented in Ref.~\onlinecite{LJBCG}
such that the  fluctuating local field at each of $^{55}$Mn
sites in  Mn$\uf$  and Mn$\ut$ ions  is originated from
the thermal excitations  in  the Zeeman-energy levels
due to the spin-phonon interaction.
Then, in order to proceed the analytical treatment,
we simplify the problem by taking into account only
the lowest two levels of the cluster spin $S$=10,
the ground-state and the first excited state
which lies above about 12\,K.
This simplification  will be reasonable,
since  the data for the $^{55}$Mn NMR  are available only below about
2.5\,K,
thus the statistical weight of higher excited states
being extremely small.
For the interpretation for $T_2$,
which was measured by observing the spin-echo decay time,
we employ  the general treatment based on non-linear theory.
As we shall explain,
it turns out that the transverse relaxation rate $T_{2}^{-1}$
is reasonably understood in terms of the strong collision regime
instead of weak collision regime.
On the other hand, the results of $T_1$ is interpreted satisfactorily
by the standard perturbation formalism.  However,
our standpoint for the interpretation for $T_1$ is
somewhat different from that presented in Ref.~\onlinecite{FWKBG}
apart from the simplification of the problem.
The essential part has already been reported
in our recent brief publication.~\cite{KGKA}

The constitution of the present article is as follows.
In the next section, we explain briefly the magnetic structure and hyperfine
coupling tensors of $^{55}$Mn in Mn$_{12}$Ac determined by analysis of
$^{55}$Mn NMR spectra.  The experimental results are presented in Sec.
\ref{sec:exp}.
Section \ref{sec:An} is devoted
to the derivation of the theoretical equations for interpretations
for the experimental results.  Subsequently we show
that the experimental results of $T_2$ and $T_1$ are interpreted
reasonably in terms of the present theoretical treatment.
In Sec.~\ref{sec:discussions} we discuss on the quantitative considerations.
 The final section yields the summaries of this article.

\section{Properties of $\mbox{Mn}_{12}\mbox{Ac}$\label{sec:system}}
\subsection{Crystal Structure and magnetic properties\label{subsec:crystal}}
\begin{figure}
\includegraphics[width=8.5cm]{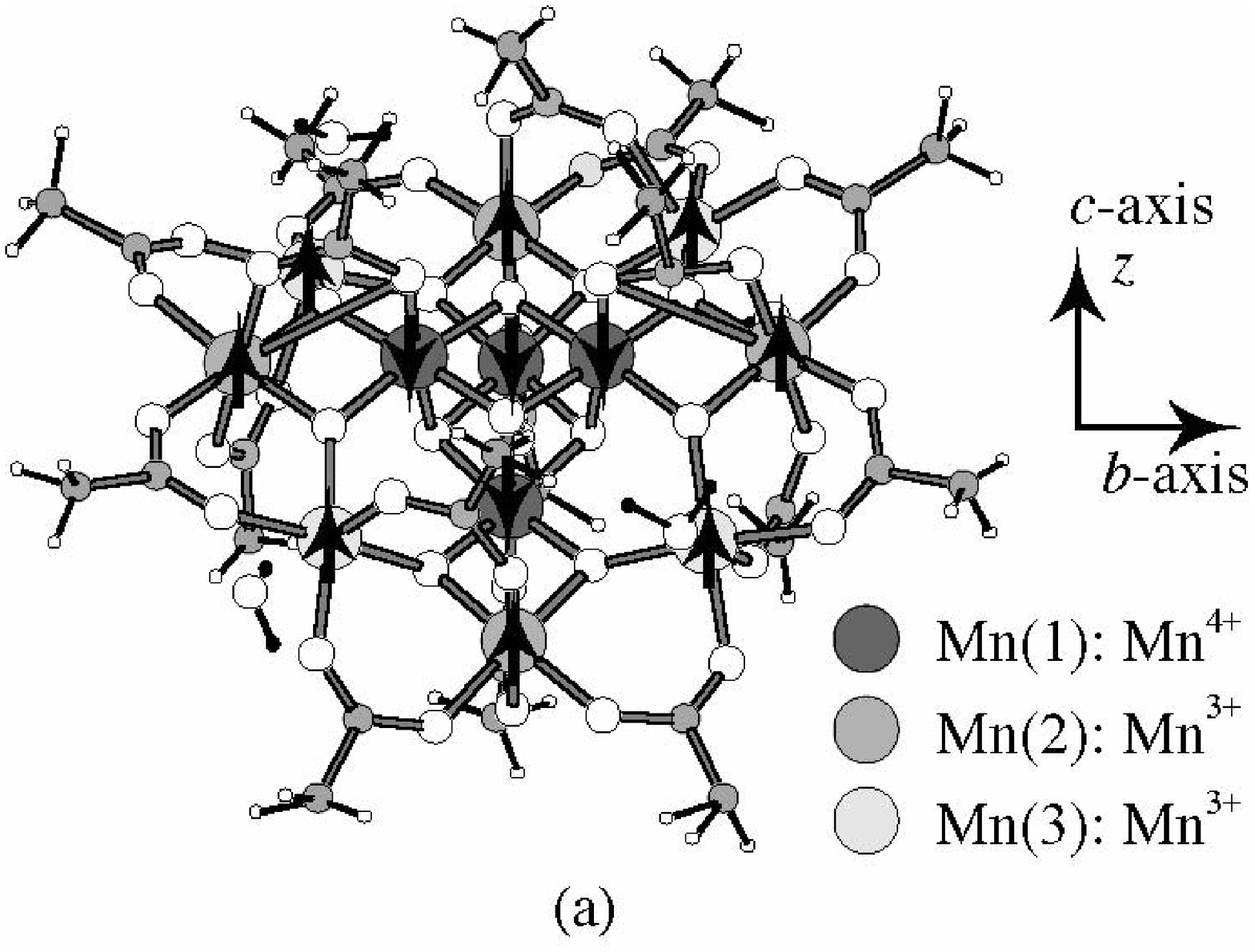}
\includegraphics[height=5.5cm]{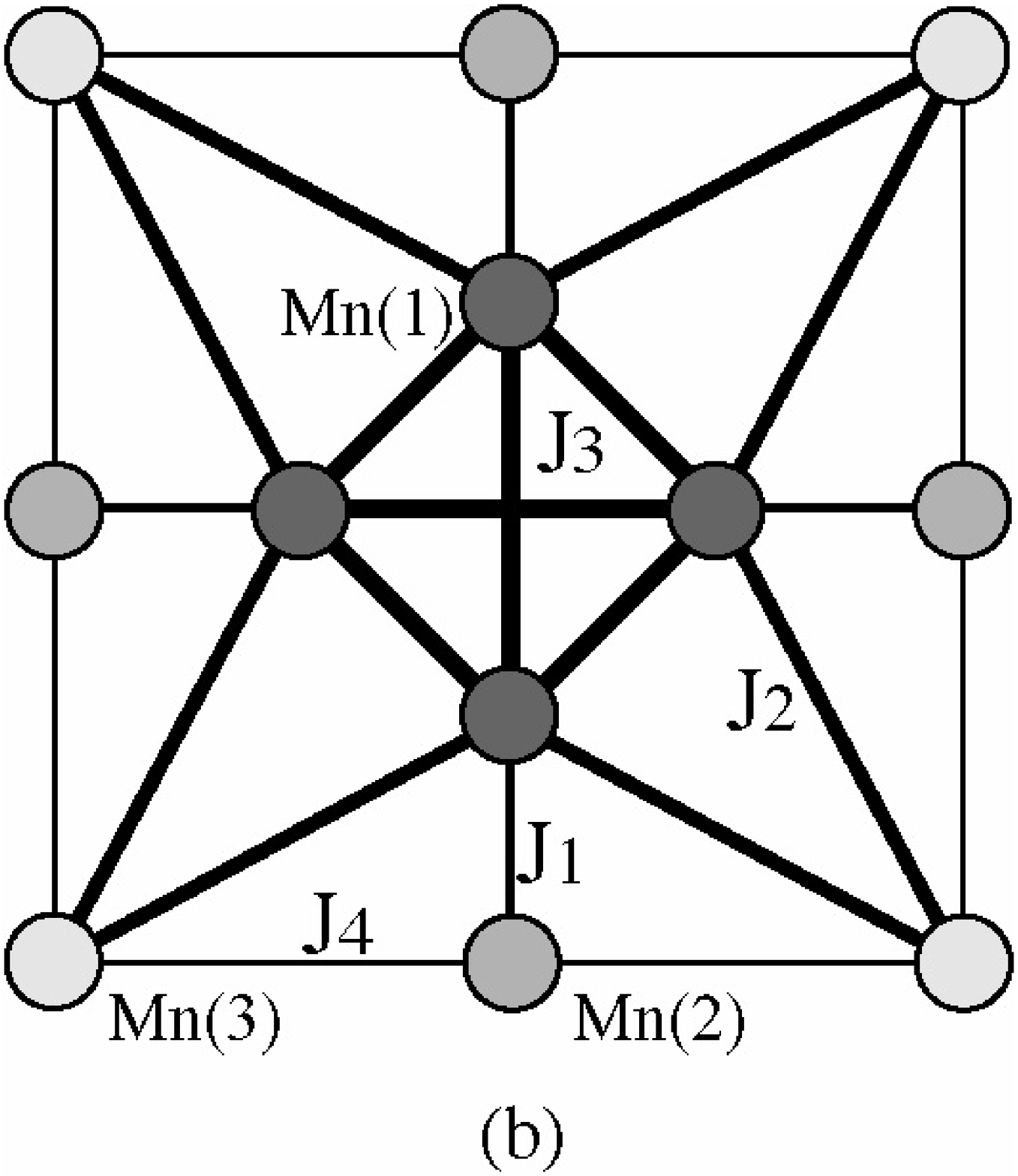}
\caption{\label{fig:crystal}
Crystal Structure of  molecular cluster
of Mn$_{12}$Ac.
The crystalline $c$-axis (the $z$-axis), along which the structure is
viewed, is taken
to tilt
upward from the the surface of the text.
The deeply  and lightly shaded large circles represent Mn$\uf$
ion and Mn$\ut$ ions, respectively.
The shaded and open circles,
and the small closed circles represent carbon, oxygen,
and proton of water molecule,
respectively.
The arrows on each of the manganese sites indicate the directions of the
magnetic moments
which are parallel or antiparallel to the $c$-axis.
(b) Schematic drawing of the exchange interactions
among the manganese ions, $J_i$ ($i=1,\cdots, 4$),
whose values are referred to the text.}
\end{figure}

The  crystal and magnetic structures of the cluster of Mn$_{12}$Ac
are shown in Fig.~\ref{fig:crystal}(a). Each cluster, which constitutes a
tetragonal  symmetry with the lattice constants of $a=b=17.3$\,\AA, and
$c=12.4$\,\AA, is constructed from four Mn$\uf$ ions ($S=3/2$) in a
central tetrahedron (denoted by Mn(1)) and surrounding eight
Mn$\ut$ ($S=3/2$) with two in-equivalent sites (denoted by Mn(2) and
Mn(3)) located alternately. The Mn atoms are linked by triply bridged
oxo oxygens and by carboxylate bridges from acetate anions. 
The Mn(1)  and Mn(2) have distorted octahedral coordination of oxygens due to
above links.  In Mn(3), one water molecule completes the octahedral
environment of oxygens.~\cite{L}


There exist four kinds of exchange interactions among these
manganese ions,
as shown schematically in Fig.~\ref{fig:crystal}(b).
Quite recently,
a set of the magnitudes and signs of these exchange couplings
has been determined, 
using  exact diagonalization of the spin Hamiltonian
of the cluster spin by a Lanzcos algorism, 
so as to explain the high-field magnetization and the realization of
the ground state of $S=10$. These values are $J_1/k_B=-119$\,K, 
$J_2/k_B=-118$\,K
(antiferromagnetic), $J_3/k_B=8$\,K (ferromagnetic) and,
$J_4/k_B=-23$\,K,~\cite{RJSGV} and except the value of $J_1/k_B$, 
they are different largely 
from the previously available values.~\cite{STSWVFGCH}  

As a combined effect of the above exchange interactions,
the ground-state configuration of the total spin of $S=10$ is
established~\cite{CGSBBG} in
such a way that the assemblies of outer eight Mn$\ut$ ions and the
inner Mn$\uf$ ions, which have respectively resultant ferromagnetic
spins of $S=16$ ($8\times2$) and $S=6$ ($4\times3/2$), are coupled
anti-ferromagnetically to each other at low
temperatures.~\cite{CGSBBG} Because of the anisotropy due to
Jahn-Teller distortion of Mn$\ut$ ions, there appears  a
single-ion type anisotropy $D$ along the $\pm c$-axis. Thus the
total magnetic moments of each cluster are either parallel or
anti-parallel to the $c$-axis.
Henceforth these clusters are
referred to respectively as \ppp and \pmp cluster, with respect to  
the $c$-axis taken as the $z$-axis.
The inter-cluster interaction is only due to dipolar origin
of the order of 0.5\,K.  In the presence of an external field $H_0$
applied along the $c$-axis, the effective exchange Hamiltonian for each
cluster is given by
\begin{equation}
{\cal H}=-DS_z^2-BS_z^4\pm g_{\parallel}\mu_B H_0S_z, \label{eq:one}
\end{equation}
where the exchange parameters  have  been evaluated from recent
high-field ESR~\cite{BGS} and neutron spectroscopy~\cite{MHCAGIC} as
$g_{\parallel}=1.93$,~\cite{BGS}
$D/k_B=0.67$\,K,~\cite{BGS}
and 0.66\,K,~\cite{MHCAGIC}
and $B/k_B=1.1$-$1.2\times10^{3}$\,K,~\cite{BGS}
and the sign $\pm$
correspond to the \ppp and \pmp clusters,
respectively. 
According to this
Hamiltonian, the discrete energy
levels $|S, m\>$ are well defined along the $z$-axis in the
ground-state spin configuration of $S=10$. 
This
gives satisfactory understandings for the most of magnetic
properties of Mn$_{12}$Ac such as an energy barrier of about 60\,K in
zero field  from  the ground level of $m=\pm10$ up to  the highest
level of $m=0$,  and 
the step-wise recovery of the magnetization at
each step of about 0.45\,T which
occurs corresponding to  the coincidence of the energy levels between
$S_z= m (<0)$ in the \ppp cluster and $S_z=m (>0)$ in the \pmp
cluster. In addition to the above main Hamiltonian, there exists
the perturbing Hamiltonian ${\cal H^{\prime}}$
including  the terms such as  higher-order transverse anisotropy
and the transverse external field. 
These terms  do not commute with $S_z$, thus playing  a  crucial role
for the mechanism of the tunnelling.
In particular, the transverse external field gives rise
to a drastic change in the tunnel-splitting
at the level-crossing fields, which may promote the tunneling
appreciably.~\cite{PS}

\subsection{Hyperfine interaction tensor\label{subsec:HIT}}
\begin{figure}
\includegraphics[width= 8.5cm]{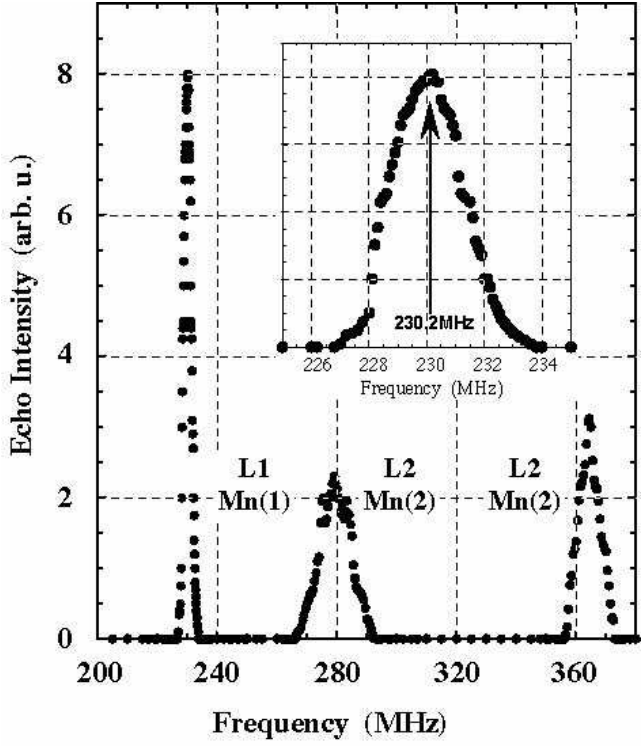}
\caption{\label{fig:spec}$^{55}$Mn NMR spectra in
Mn$_{12}$Ac obtained at 1.4\,K in zero field.
The three resonance lines labeled by L1, L2, and L3 are
associated with Mn$\uf$ (Mn(1)), and two inequivalent 
Mn$\ut$ ions (Mn(2) and Mn(3)), respectively.
The inset shows the detail of the L1 line.}
\end{figure}

Next we review the $^{55}$Mn
hyperfine interaction tensor determined 
from the
NMR spectra in zero
field in Ref.~\onlinecite{KGKTA}
for the numerical evaluation
of the present experimental results.
Figure~\ref{fig:spec} shows the NMR spectra of $^{55}$Mn 
($I=5/2$) in Mn$_{12}$Ac obtained at 1.45K in zero field.    
The three completely separated lines with the central
frequencies of $\nu_N=$230, 279, and 364\,MHz, 
were identified to be due to Mn$\uf$ ion (Mn(1)), and Mn$^{3+}$ ion (Mn(2) and Mn(3)), respectively.
The corresponding internal fields are  
$H_{{\rm int}}=21.8, 26.5, 34.5$\, T.
From now on, these three lines are referred to as L1, L2, and L3, respectively.
The resonance lines L1,
L2, and L3  involve the five-fold quadrupole-splitting  
with $\Delta\nu_q = 0.72, 4.3, 2.9$\,MHz, 
respectively.~\cite{KGKTA}

The nuclear hyperfine Hamiltonian, which consists of Fermi-contact, dipolar, and orbital terms, 
is obtained by taking the expectation values
for the corresponding Hamiltonians with respect
to the ground-state wave-function of the magnetic ion. This Hamiltonian is expressed as,  
\begin{equation}
{\cal H}_N={\bf I}\cdot A\cdot{\bf S}
=-\gamma_N\hbar{\bf I}\cdot({\bf H}_F+{\bf H}_d+{\bf H}_{l}),
\label{eq:two}
\end{equation}
where $A$ is the hyperfine coupling tensor between
the nuclear spin
${\bf
I}$ and the electronic spin ${\bf S}$, and ${\bf H}_F$, ${\bf H}_d$, ${\bf
H}_{l}$
are the corresponding hyperfine fields. 
Each of manganese ions in Mn$_{12}$Ac
is subject to the crystalline field with dominant cubic symmetry
due to the surrounding distorted octahedral coordination of oxygens. 

The ground-state of the Mn$^{4+}$ ion (3d$^3$, $^4$F) is the orbital-singlet.
So the dipolar and the orbital terms in Eq.~(\ref{eq:two}) vanishes primarily, and
only the isotropic Fermi-contact term is important.  
The corresponding Fermi-contact field ${\bf H}_F$ is given, using 
the conversion factor of 1 atomic unit (\,$a.u.$) = 4.17\,T, as
\begin{equation}
 {\bf H}_F=-\frac{A_f{\bf S}}{\gamma_N\hbar}=-2\times4.17\chi {\bf S},
\label{eq:three}
\end{equation}
where $A_f$ is the component of the tensor $A$, and
$\chi$ is the effective field per unpaired electron in atomic unit. 
According to 
 Freeman and Watson,~\cite{FW} 
the value of $\chi$ is calculated to be 2.34 for the free Mn$\uf$ ion,
by taking into account the contributions from three
inner electron spins (1$s$, 2$s$, and 3$s$).
The minus sign in ${\bf H}_F$ indicates that the direction of ${\bf H}_F$ is
opposite to the magnetic moment. Thus for the free Mn$\uf$ ion, 
the values of ${\bf H}_F$ is estimated to be $H_F=29.3$\,T.
The effect of the admixture of the higher
triplet state to ground-state and 
distorted crystalline field
 may be extremely small.
In fact, for the diluted ion in the trigonal
symmetry, the hyperfine anisotropy is of the order of 0.1\%.~\cite{FW}.
It is noted that the experimental
internal field of 21.8\,T 
is smaller by 26\% as compared with calculated value of  29.3\,T for
the free ion. This is understood by considering 
a large amount of reduction of the magnetic moment.
The recent polarized neutron diffraction measurement
yields 
the presence of 22\% reduction from the full value of
3$\mu_B$ in the magnitude of the Mn$\uf$ magnetic
moment.~\cite{RBAHA} Since  the reduction of the magnetic moment due
to 3$d$-electrons reflects the contact term, 
the present result is consistent
with the neutron results. 
Such a large amount of reduction may be
ascribed to the presence of the covalence and strong exchange
interactions, as observed usually in the condensed matter. 

On the other hand, the ground state of the Mn$\ut$ ion (3d$^4$, $^5$D) is 
orbital doublet denoted by $E_g$ in the cubic crystalline field.
As in the case of Mn$^{4+}$ ion, the Fermi-contact field $H-F$  
is obtained for the free Mn$^{3+}$ to be 48.5\,T 
using the calculated value of $\chi$=2.91.~\cite{FW}
Because of the additional
elongated tetragonal symmetry of the crystalline field caused by
Jahn-Teller effect and low-symmetric carboxylate ligands, the orbital
degeneracy is removed to the lower and higher states with
the  wave functions expressed as 
$|\Psi_1\>=|X^2-Y^2\>$ and  $|\Psi_2\>=|3Z^2-r^2\>$, respectively.
Here we defined the rectangular coordinate system ($XYZ$) with the
tetragonal $Z$-axis and the $X$-axis along one of the principal axes
in the tetragonal plane. 
Further, the orthorhombic
distortion of the crystalline field gives rise to the admixture of $|\Psi_2\>$ to
$|\Psi_1\>$, thus  the ground-state wave-function being 
expressed as
$|\Psi_g\>= \cos\phi|\Psi_1\> +\sin\phi|\Psi_2\>$,
where $\cos\phi$ and $\sin\phi$ are 
normalization factors.
In Mn(2) and Mn(3), the
$Z$-axis tilts from the $c(z)$-axis by the angles of $\theta=11.7^\circ$
and $36.2^\circ$, respectively.
In the case of Mn$\ut$ ion, 
the dipolar term in Eq.~\ref{eq:two} contributes 
appreciably to the hyperfine tensor in addition to the dominant isotropic Fermi-contact term. 
By using  the wave function $|\Psi_g\>$ in  the standard
operator-equivalent method, the dipolar Hamiltonian
${\cal H}_d$=$-\gamma_N\hbar{\bf I}$$\cdot$${\bf H}_d$, which is 
expressed in terms of the principal terms 
with respect to the ($XYZ$) coordinate system, is given as,   
\begin{equation}
 {\cal H}_d=h_d\gamma_N\hbar\cos(2\phi)
     \left[S_ZI_Z-\frac{1}{2}(S_XI_X+S_YI_Y)\right]
\label{eq:four}
\end{equation}
with
\begin{equation*}
 h_d=\frac{4}{7}\mu_B\<r^{-3}\>_d,
\end{equation*}
where $\<r^{-3}\>_d$  is the average for the orbital radius $r$ with
respect to 3$d$ shell. It is useful to define the
($xyz$) rectangular coordinate system with the $z$-axis along the magnetic
moment
and the $x$-axis is taken in the $Zz$-plane.
(See Fig. 4 in Ref.~\onlinecite{KGKTA}.)
Then the above equation is expressed as~\cite{KGKTA}
\begin{equation}
 {\cal H}_d={\bf I}\cdot D\cdot{\bf S}\nonumber
\end{equation}
with
\begin{equation}
D=\frac{h_d}{4}\gamma_N\hbar\cos(2\phi)
\begin{pmatrix}
       2(2-3\cos^2\theta)&0&-3\sin2\theta\\
      0&-2&0\\
     -3\sin2\theta&0&2(3\cos^2\theta-1)
\end{pmatrix}.
\label{eq:five}
\end{equation}
The principal terms are anisotropic, and there appear the
off-diagonal terms.
For the free Mn$\ut$ ion which has
$\<r^{-3}\>_d=4.8\,
a.u.$,~\cite{FW} $h_d$ is evaluated to be $+$17.1T.

The orbital contribution is in general evaluated
from the
equation,
$H_{l}=-2\mu_B\<r^{-3}\>_{d}\Delta g$, where $\Delta g$
represents the deviation of the $g$-value from $g_s=2.0023$ due to
admixture of the higher excited orbital state. Applying the value of
$g_z =1.95$ and $g_{\perp} =1.97$ for Mn$\ut$ ion evaluated from
high field EPR measurement,~\cite{BGS} the value of $H_d$  is
calculated, using  $\<r^{-3}\>_d=4.8\,a.u.$ for the free ion, to be
about 2\,T. 
So, 
we may neglect the orbital
contribution as compared with the other terms.
By taking into account the dipolar contribution to
the internal field,  which is given by the $D_{zz}$ component in the
dipolar tensor $D$ (Eq.~\eqref{eq:five}),
the total internal field is given by
\begin{equation}
 H_{{\rm int}}=|{\bf H}_F|-D_{zz}/\gamma_N\hbar.
\label{eq:six}
\end{equation}
The identification of the L2 and
L3 lines was made in view of Eq. (\ref{eq:six}).
By considering that the increase in $\theta$
plays a role to
deduce
$|{\bf H}_F|$ 
as far as $\theta < 53^\circ$,
it turns out that  the line L2 with lower $\omega_N$  
should be due to Mn(2), and thus the L3 line  being due to Mn(3).
As for the quadrupole splitting, the dominant term  
is expressed, for the axial-symmetry case, as
$
\Delta\nu_q=\frac{1}{2}(3\cos^2\theta-1)\nu_q,
$
where $\nu_q$ is the quadrupole-splitting parameter.
The larger value of $\theta$ yields the larger
splitting as far as $\theta < 53^\circ$. 
The above identification is then consistent with the
observed difference in quadrupole splitting between 
$^{55}$Mn NMR for the L2 and L3 lines.

The determination of the components of the hyperfine tensors $A$ for  
Mn(2) and Mn(3) of Mn$\ut$ ion was made in the following way.
There are three
unknown factors, the reduction factor for the contact term, 
the value of $\<r^{-3}\>_d$, and the amount of the mixing of the two wave
functions in the ground-state $E_g$, which is represented by the factor
$\cos2\phi$ in Eq.~(\ref{eq:five}). 
According to Ref.~\onlinecite{RBAHA}, the reductions 
of the magnetic moments for the  Mn(2) and Mn(3) ions are obtained 
to be 8\% and 6\%, respectively. 
First, by applying  these reduction factors 
to the value of $\<r^{-3}\>_d$ in
Eq.~\eqref{eq:four} for each ion, we evaluated that $h_d$=$+$15.7 and $+$ 16.0\,T, respectively.
Secondly, referring to the crystal parameters given in Ref.~\onlinecite{L}, we
assumed the tetragonal and orthorhombic symmetries
of the crystalline field for Mn(2) and Mn(3), respectively. 
That is, we put $\cos2\phi=1$ for Mn(2), and the coefficient $\cos2\phi$ for Mn(3) 
was remained as an unknown factor.  
Then using
$\theta =11^\circ$ for Mn(2), we obtain the dipolar contribution
$D_{zz}/\gamma_N\hbar=$+$1.6$\,T. Using the experimental value of
26.5\,T for $H_{{\rm int}}$ in Eq.~\eqref{eq:six},  we find $H_F$(Mn(2))=24.8\,T.
This corresponds to 85\% of $H_{{\rm int}}$ for the free ion, thus reduction
factor for the contact term being  evaluated to be 15\%.
Next we  adopted the
same value of the contact field for Mn(3).  
Then using $\theta=36^\circ$, 
the mixing parameter for Mn(3) was estimated to be $\cos2\phi=0.89$.  
From the above considerations, we finally determined  the following numerical values of
the components of the $^{55}$Mn hyperfine-interaction tensors for each of
three manganese ions, which are expressed in unit of MHz with respect to
the ($xyz$)-coordinate frame with the $z$-axis of the
$c$-axis;
\begin{subequations}\label{eq:Atensor}
\begin{align}
A({\rm Mn(1) })&={\rm diag}(153,153,153),\\
A({\rm Mn(2)})&=\begin{pmatrix}
    254&0&-24.7\\
  0&176&0\\
 -24.7&0&140
\end{pmatrix},\\
A({\rm Mn(3)})&=\begin{pmatrix}
    221&0&-53.0\\
  0&181&0\\
 -53.0&0&182
\end{pmatrix}.
\end{align}
\end{subequations}

\section{Experimental results for the relaxation rates\label{sec:exp}}
\begin{figure}
\includegraphics[width= 8.5cm]{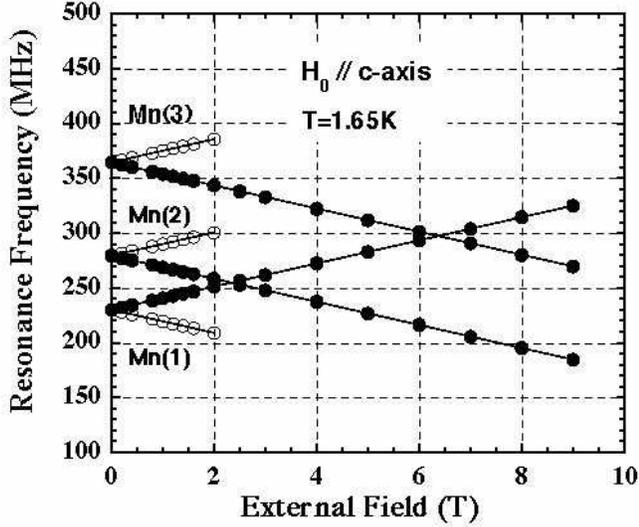}
\caption{\label{fig:specfield}
The field dependence of the resonance frequencies  of the
central peak of  the three resonance lines L1, L2, and L3 
obtained at 1.65\,K.
The
external field is applied along the $c$-axis. 
The closed and open
circles correspond to the $^{55}$Mn NMR belonging to 
the \ppp and \pmp clusters, respectively.}
\end{figure}

The transverse and longitudinal relaxation times $T_2$ and $T_1$ of
the $^{55}$Mn were measured  for the three resonance lines at the
liq.Helium temperatures with the external field $H_0$ up to 9\,T applied
along the $c$-axis,
and for the L1 line the measurement was extended
down to 200\,mK  in zero field.  Figure~\ref{fig:specfield} shows the
field dependence of the resonance frequencies of the central peaks of
the three resonance lines, which was obtained at 1.65K
by applying the external field $H_0$ along the $c$-axis ($z$-axis)
after zero-field cooling.
Within the experimental error, the slopes of these lines
coincide with the gyromagnetic ratio $\gamma_N$ of the free manganese nuclei.
As explained in Sec. \ref{subsec:HIT}, the internal field
$H_{{\rm int}}$ of $^{55}$Mn, which is mainly due to the Fermi-contact term,
appears along the spin direction, that is, opposite to the magnetic
moment. So when $H_{{\rm int}}\gg H_0$ as in the present case,  the
resonance conditions for Mn$\uf$ ion belonging to
the \ppp and \pmp clusters are given by
$\omega_N=\gamma_N(H_{{\rm int}}\pm H_0)$, and these are referred
to as upper and lower branches, respectively.
 In the case of  Mn$^{3+}$ ion, the above relations are
vice versa.

\begin{figure}
\includegraphics[width=8.5cm]{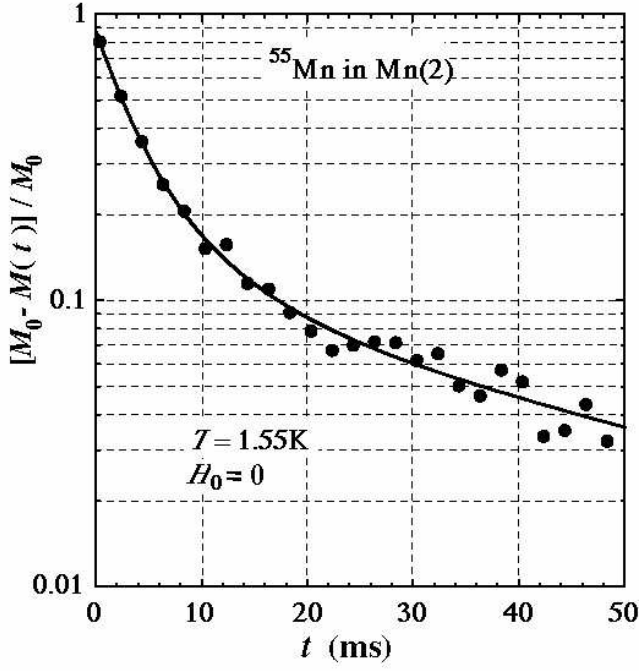}
\caption{\label{fig:T1fitting}
A typical example of the recovery of the $^{55}$Mn nuclear magnetization $M(t)$
measured for Mn(2) at $T=1.55$\,K in zero field
as a function of the time $t$ between  the end of the saturation $rf$-pulse
and the beginning of the searching $rf$-pulse,
which is normalized with the equilibrium value $M_0$.
This is a fitting curve shown by the solid line
yields $T_1=44.7$\,ms using the best fit values of the
coefficients $a=0.11$, $b=0.24$, and $c=0.51$ in Eq.~(\ref{eq:T1recovery}).}
\end{figure}

At lower temperatures than the blocking temperature of
$T_B\approx3$\,K, we can observe the NMR signals and measure
$T_2$ and $T_1$ for the \pmp
cluster in addition to those for the \ppp cluster, as far as  the
relaxation time of the reorientation of the \pmp clusters to the
$z$-direction is enough longer as compared with $T_1$. 
The transverse relaxation time $T_2$
were obtained by measuring the decay of spin-echo amplitude 
as a function of the time interval between two $rf$-pulses.
The decay was of single-exponential type.
The longitudinal relaxation time $T_1$ is obtained in general by measuring 
the recovery of the nuclear magnetization after the saturation of the
nuclear magnetization of the central line.  Under the ideal condition
of the complete saturation, the magnetization recovery for the
nucleus of  $I=5/2$ is obtained by the equation~\cite{AT}
\begin{align}\label{eq:T1recovery}
m(t)&=1-\frac{M(t)}{M_0}\nonumber\\
&=a\exp(-t/T_1)+b\exp(-6t/T_1)+c\exp(-15t/T_1)
\end{align}
with the condition such that $a+b+c=1$. 
However, in the present case, it was difficult to attain complete saturation
of the NMR signal  
because of the
broadness of each quadrupole splitting resonance line.
Then the
relaxation rate $T_1$ was determined by 
doing the best-fitting of the experimental recovery curve to the 
Eq.~(\ref{eq:T1recovery}).
The value of
$T_1$ obtained in such a way was almost the same as the value
determined from the fitting of the slowest recovery region to the
single exponential equation $\exp(-t/T_1)$.
Figure ~\ref{fig:T1fitting} shows a typical example of the best-fit recovery
curve of the nuclear magnetization for the Mn$\uf$ ion.

\begin{figure}
\includegraphics[width= 8.5cm]{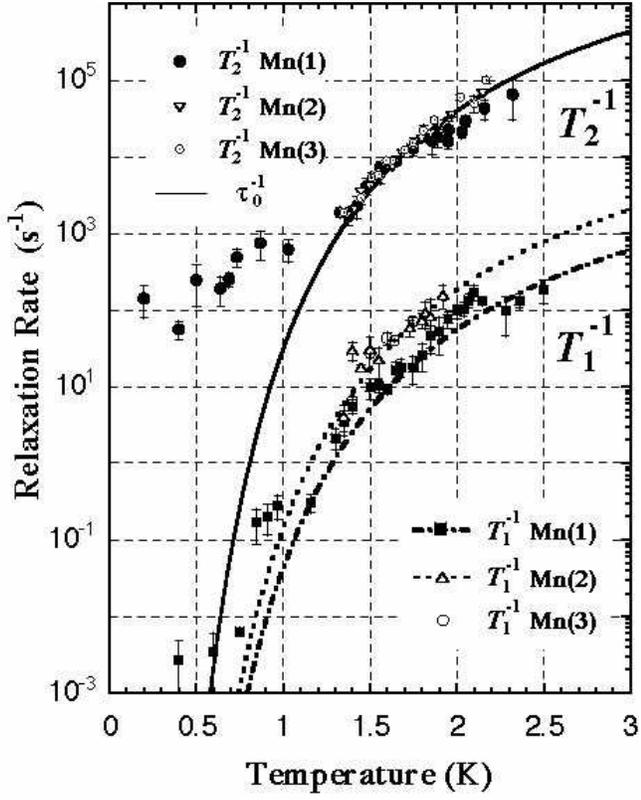}
\caption{\label{fig:Tdep}
Temperature dependence of $T_{2}^{-1}$ and
$T_{1}^{-1}$ of $^{55}$Mn  for Mn$\uf$ ion (Mn(1)) and
Mn$\ut$ ions (Mn(2) and Mn(3)) in zero field.
The solid line represents the temperature dependence
of $\tau_0^{-1}$ (Eq. (\ref{eq:T2-B})).
The dot-dashed and dotted lines represent
the best fit of Eq. (\ref{eq:T1-B}) to 
the experimental results for Mn$\uf$ ion and 
for Mn$\ut$ ion (Mn(2)), respectively.
The temperature dependence of Eq. (\ref{eq:T1-B})
is given by $\tau_0^{-1}$.}
\end{figure}

Figure.~\ref{fig:Tdep} show the temperature dependence of 
the transverse relaxation rate $T_{2}^{-1}$
and the longitudinal relaxation rate 
$T_{1}^{-1}$ measured in
zero field for each central peak of the three resonance lines.
As is seen, both rates exhibit qualitatively  the same  remarkable decrease
with decreasing temperature above about 1.4\,K,
and the values of $T_{1}^{-1}$ are smaller than that of $T_{2}^{-1}$ by
almost two orders of magnitudes.
Below 1.4\,K,  $T_{2}^{-1}$ becomes rather moderate, and it becomes almost
constant around the value of 100\,s$^{-1}$ below about 0.5\,K.
The values of $T_{2}^{-1}$ for Mn$\uf$ ion and Mn$\ut$
ions are almost the same, though the former is somewhat smaller only
at the temperatures above about 2\,K.
On the other hand, the value of $T_{1}^{-1}$ continues to decrease
remarkably down
to around 0.5\,K,  Thus, at very low temperatures there appears, between 
$T_{2}^{-1}$ and $T_{1}^{-1}$, 
a difference extending over four orders of magnitude.
The values of $T_{1}^{-1}$  for Mn$^{3+}$ ion are larger
by about twice than that for Mn$^{4+}$ ion.

The field dependence of $T_{2}^{-1}$  for 
Mn(1), Mn(2), and Mn(3).
Mn$\uf$ ion was
measured at 1.65\,K for the \ppp cluster (upper branch) up to 9\,T
and for the \pmp cluster (lower branch) up to 1.2T.
The data were taken also at 1.45\,K only for the \ppp cluster up to 9\,T.
The field dependence of $T_{1}^{-1}$ for Mn$^{4+}$ ion was measured at
1.65\,K for the \ppp cluster (upper branch) up to 5\,T, and for the \pmp
cluster (lower branch) up to 1.2\,T. The data were taken also at 1.45\,K 
for the \ppp cluster up to 3\,T.
The experimental results are shown in Fig.~\ref{fig:Fdep} and
~\ref{fig:branch}.
As is seen, $T_{2}^{-1}$ for the \ppp cluster (upper branch)
decrease monotonously with increasing field down to the field at which
$T_{2}^{-1}$
reaches the value of around 150 \,s$^{-1}$. This value is nearly close
to the constant value obtained at very low temperature in  the
temperature dependence of $T_{2}^{-1}$.  The field  dependence  at 1.45\,K
is slightly remarkable  than that for 1.65\,K.
The anomalous  peak around $H_0=6.8$\,T in $T_{2}^{-1}$ may be
due to cross-relaxation  with $^1$H NMR. It should be noted that the field
dependence of $T_{1}^{-1}$ is slightly more remarkable than that of
$T_{2}^{-1}$.  For the \pmp
cluster (lower branch), on the other hand, the values of $T_{2}^{-1}$ increase
with increasing field. However, the change is rather monotonous as in the
case of
the \ppp cluster. No any appreciable change was observed at the level-crossing
fields around $H_0$=0, 0.45, and 0.9T.
Figure~\ref{fig:compFdep} represents the field dependence
 of $T_{2}^{-1}$ for Mn$\uf$ ion (Mn(1)) 
and for Mn$\ut$ ion (Mn(2))   
for the \ppp cluster obtained at 1.45K.  No appreciable difference was found.


\section{Analysis\label{sec:An}}
In this section, we shall analyze the  experimental results.
The nuclear magnetic relaxations in Mn$_{12}$Ac will be primarily
caused by the fluctuating component $\delta {\bf S}(t)$ of the on-site
manganese ion via the hyperfine interaction
$\delta{\cal H}_N(t)={\bf I}\cdot A\cdot\delta {\bf S}(t)$,
where $A$ is the hyperfine interaction tensor given by Eq.~\ref{eq:Atensor}.
In view of the fact that an
assembly of the strongly-coupled manganese spins in a cluster is
established as the cluster spin of $S=10$, we may assume that each
manganese spin is subject to the same fluctuation corresponding to
thermal fluctuation of the cluster spin along the z-axis,
which is caused by
the spin-phonon interaction.
Then, the effective perturbing hyperfine interaction is expressed as
\begin{equation}
\delta{\cal H}_N(t)=(I_xA_{xz}+I_yA_{yz}+I_zA_{zz})\delta S_z(t).
\end{equation}
Lascialfari \etal. treated the fluctuating local field
due to the spin-phonon interaction in analyzing the proton and muon
spin-lattice relaxation rates in Mn$_{12}$Ac
by considering all of Zeeman levels with statistical weight.~\cite{LJBCG}
Here, we simplify the problem by taking into account only the lowest
two energy-levels of $S=10$ within  
each of the double-well potential, that is,  
the ground-state $S_z=m=-10$ and the
first excited state $m=-9$ 
for the (+) cluster, and $m=+10$ and $m=+9$ for the (-) cluster. 
Such a simplification is reasonable since the present
$^{55}$Mn NMR is available only at low temperatures below about 2.5\,K where the
statistical weights of the higher excited states are quite small.
Then the average life-times $\tau_0$ and $\tau_1$ of
the ground-state and the excited state are given as follows
within the framework of the lowest two levels;
\begin{subequations}\label{eq:ratio}
\begin{equation}
\frac{1}{\tau_0}=\frac{C\Delta^{3}}{\exp (\Delta/T)-1}
\end{equation}
and
\begin{equation}
\frac{1}{\tau_1}=\frac{C\Delta^{3}}{1-\exp (-\Delta/T)},
\end{equation}
\end{subequations}
where $C$ is the coupling constant for the spin-phonon interaction, \cite{VBSR}
and $\Delta$ is the energy difference between the ground-state
and the first excited state, which is given by
\begin{equation}
\Delta=19D_I/k_B \pm g_{\parallel}\mu_BH_0/k_B.{\nonumber}
\end{equation}
Here the signs $\pm$
correspond to
the upper and lower branches for Mn$\uf$ ion, respectively, and 
 vice versa 
for Mn$\ut$ ion.
By using the values of
$D/k_B=0.67$\,K,~\cite{BGS}
and $g_{\parallel}=1.93$,~\cite{BGS}
$\Delta$ is given as $(12.7\pm 1.30\times H_0)$\,K,
$H_0$ being expressed in T.
In our present experimental condition for the applied field, the attained
value of $\Delta$ is
at the least about 11\,K. 
So, for the low temperatures below 2.5\,K, it turns out,
from the expressions given by Eqs.~(\ref{eq:ratio}), that
$\tau_0$ exhibits very remarkable temperature dependence, whereas $\tau_1$
remains
almost constant with
the relation such that $\tau_0\gg\tau_1$.
Then, the effective
fluctuation at each of the $^{55}$Mn sites is regarded to be
step-wise, and it is characterized by random sudden jumps
between the ground-sate and the excited state, as shown schematically
in Fig.~\ref{fig:fluc}.
Here $h_{\alpha}$ ($\alpha=z$ or $\perp$) is
the average magnitude of the effective fluctuating field along or
perpendicular to the $z$-axis. These effective fluctuating fields are
related to the components of the hyperfine interaction tensor $A$ as
$h_z=-A_{zz}/\gamma_N \hbar$ and $h_{\perp}=-A_{xz}/\gamma \hbar$.
In the followings we find expressions for the nuclear magnetic relaxation
rates on the basis of the above model to look at the experimental results.

\subsection{Transverse relaxation rate\label{subsec:Analsys T2}}

First we consider the transverse relaxation rate. In our
experiment, the relaxation time $T_2$ was determined by measuring the
time constant of the decay of the spin-echo amplitude $E(2\tau)$,
that is, the macroscopic transverse nuclear magnetic moment, as a
function of time interval $\tau$ between the two \emph{rf}-pulses.
This
decay, which corresponds to the phase disturbance of the Larmor
precessions caused by the longitudinal fluctuating local field, is
represented as~\cite{KA}
 \begin{equation}
E(2\tau)=E_0\left< \exp \bigl[i\int_0^\tau \delta\omega(t)dt-i\int
_\tau^{2\tau}
\delta\omega(t)dt]\right>
\label{eq:AAA}
\end{equation}
The spin-echo amplitude can be calculated from Eq.~\eqref{eq:AAA}
by considering all possible pulse sequences of e fluctuations,
the phase deviations and the statistical weight of the pulse sequence.
This  problem
has been treated by
Kohmoto \etal. for the interpretation of $^{133}$Cs relaxation times
$T_2$  and $T_1$ in the $S=1/2$ Ising-like linear chain
antiferromagnet CsCoCl$_3$.\cite{KGMFFKM}
According to the procedure
presented in Ref.~\onlinecite{KGMFFKM}, we obtain, for $\tau_0\gg\tau_1$, the
following final expression
\begin{equation*}
E(2\tau)=E_0\exp(-\frac{2\tau}{T_2})
\end{equation*}
with
\begin{equation}
\frac{1}{T_2}=\frac{1}{\tau_0}\cdot
\frac{(\gamma_Nh_z\tau_1)^2}{1+(\gamma_Nh_z\tau_1)^2}.
\end{equation}
Thus the relaxation rate depends on the number of the
fluctuation pulse per second, $\tau_0^{-1}$ and average phase change
$\gamma_N h_z$ for one fluctuating field.\
The above equation yields for $\gamma_Nh_z\tau_1\ll1$
\begin{subequations}\label{eq:T2}
\begin{equation}
\frac{1}{T_2}=\frac{1}{\tau_0}(\gamma_Nh_z\tau_1)^2\sim
\frac{\tau_1^2}{\tau_0}, \label{eq:T2-A}
\end{equation}
and for $\gamma_Nh_z\tau_1\gg1$
\begin{equation}
\frac{1}{T_2}=\frac{1}{\tau_0}.\label{eq:T2-B}
\end{equation}
\end{subequations}

\begin{figure}
\includegraphics[width=8.5cm]{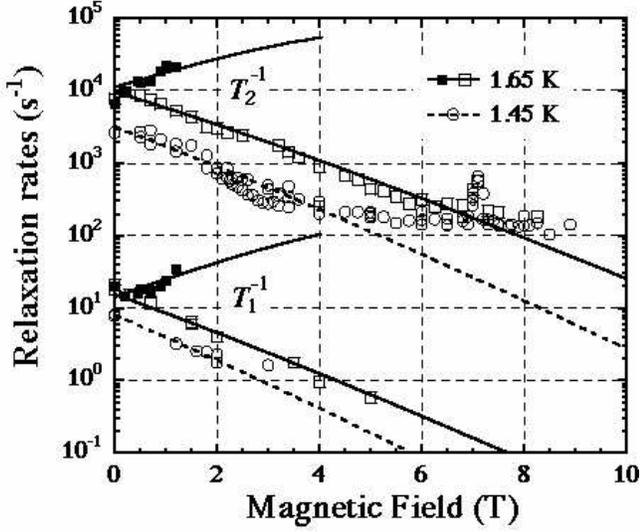}
\caption{\label{fig:Fdep}Field dependence of $T_{2}^{-1}$ and $T_{1}^{-1}$ for
the L1 line (Mn$\uf$ ion).
The external field is applied along the $c$-axis.
The open and closed squares represent the experimental results
obtained at 1.65\,K for the upper branch (\ppp cluster) 
and the lower branch (\pmp cluster), respectively.
The open circles represent the experimental results for
the upper branch obtained at 1.45\,K.
The solid and dashed lines drawn for the results of $T_2^{-1}$
represent the  theoretical equations of 
$\tau_0^{-1}$ corresponding at 1.65 and 1.45\,K,respectively.
Those for the results of  $T_1^{-1}$ represent
the best-fit of the theoretical equations of 
$\tau_0^{-1}\omega_N^{-2}$ to the experimental results.
}
\end{figure}

\begin{figure}
\includegraphics[width=8.5cm]{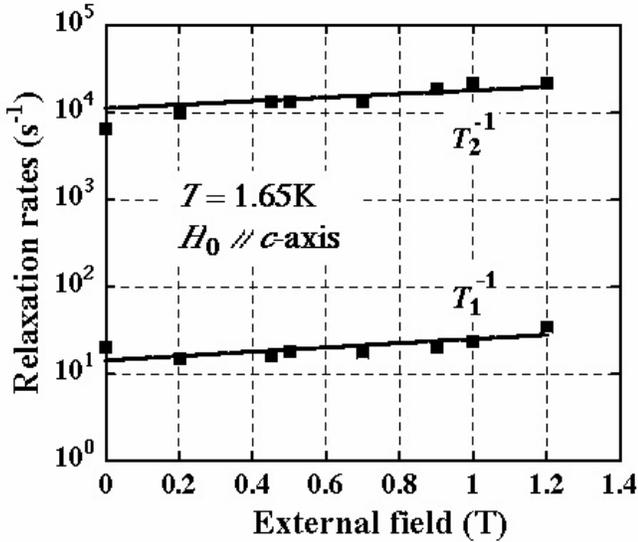}
\caption{\label{fig:branch}Field dependence of 
the relaxation rates 
$T_2^{-1}$ and $T_1^{-1}$ of
$^{55}$Mn in Mn$\uf$ ion for the the \pmp
cluster (lower branch) measured at 1.65\,K.}
\end{figure}

\begin{figure}
\includegraphics[width=8.5cm]{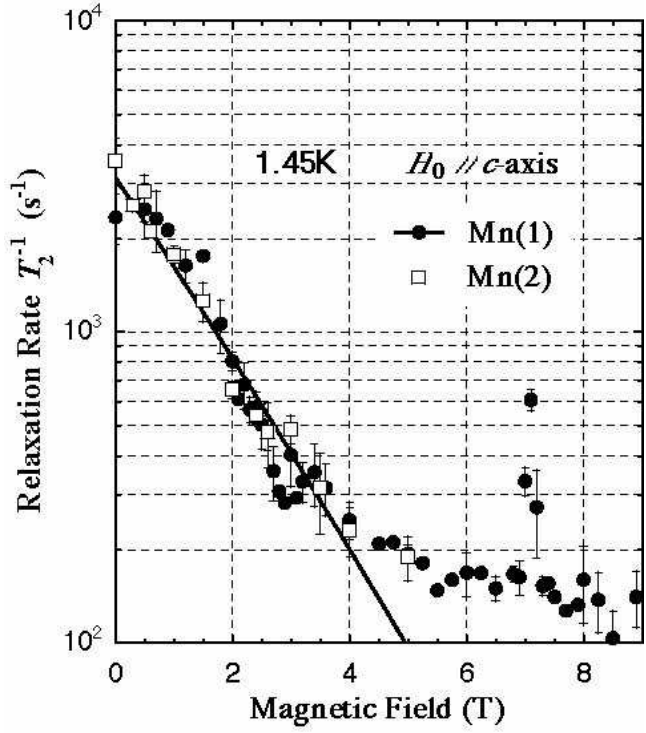}
\caption{\label{fig:compFdep}Field dependence of $T_{2}^{-1}$ of the
$^{55}$Mn in Mn(1) and Mn(2) for 
the\ppp cluster obtained at 1.45\,K.}
\end{figure}

\begin{figure}
\includegraphics[width=8.5cm]{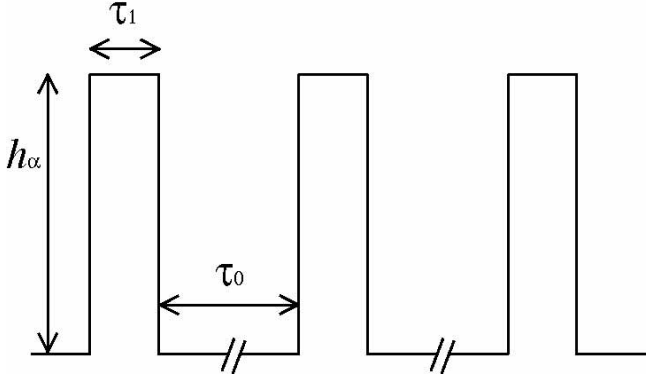}
\caption{\label{fig:fluc}Schematic drawing of the step-wise
fluctuating local field associated with the excitation from the ground-state to the first
excited state with respective average life-times of
$\tau_0$ and $\tau_1$. $h_\alpha$ represents the fluctuating field
longitudinal ($\alpha=z$) or transverse ($\alpha =\perp$) with
respect to the nuclear quantization axis, which coincides
to the $c(z)$-axis.}
\end{figure}

\begin{figure}
\includegraphics[width=8.5cm]{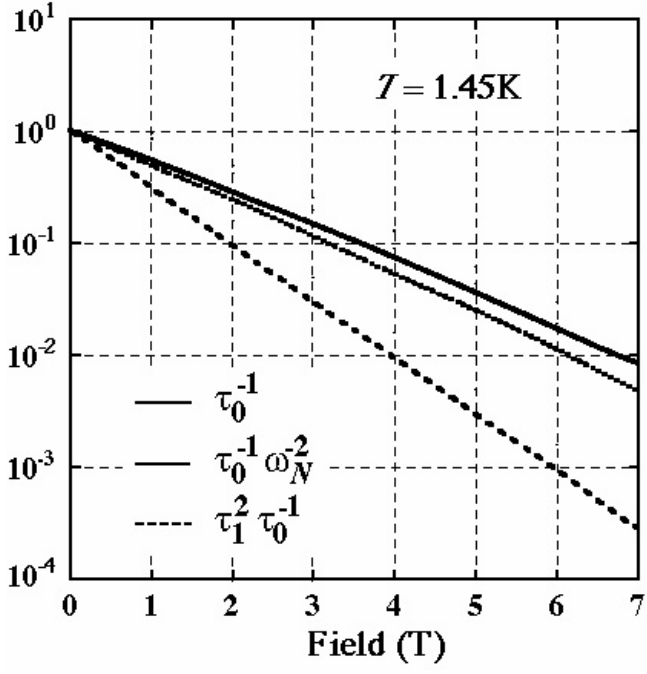}
\caption{\label{fig:rates-Hdep}
The qualitative field dependence of the relevant terms in
the theoretical equations of the relaxation rates $T_2^{-1}$ and $T_1^{-1}$ calculated for $T=1.45$\,K. 
The solid and dashed lines represent the field dependence
of $T_2^{-1}$ corresponding to the strong- and week-collision
regimes, respectively.
The dotted and dashed lines represent the field dependence of
$T_1^{-1}$ corresponding to
the high- and low-frequency limits, respectively.
These curves are normalized at $H_0=0$.
}
\end{figure}

As we see in the following section, Eq.~\eqref{eq:T2-A} corresponds to the
expression obtained on the basis of the standard perturbation method,
that is, the weak collision regime.
On the other hand, Eq.~\eqref{eq:T2-B}
means that the relaxation rate is
determined solely by the average number of the appearance
of the thermal excitation per one second, and further it does not depend
on the magnitude of the fluctuating field.
This is the strong collision regime.
As evaluated above, in the present experimental conditions, the temperature
dependence of $T_{2}^{-1}$ results predominantly through the term of
$\tau_0$.
Accordingly, as far as the temperature dependence
is concerned, there is no appreciable difference
between both regimes for $T_{2}^{-1}$.
On the other hand, since the
term of $\Delta^3$ in $\tau_0$ and $\tau_1$ contributes to
the field dependence of $T_{2}^{-1}$,
the qualitative difference between Eqs.~\eqref{eq:T2-A} and ~\eqref{eq:T2-B}
is expected to appear in
the field dependence of $T_{2}^{-1}$.
Thus it is worthwhile
to look at the qualitative field dependence of $T_2$ to find the plausible
regime.
The field-dependence of the relevant terms of  $\tau_0$ and
$\tau_1^2 \tau_0^{-1}$ calculated for T=1.45\,K are shown in
Fig.~\ref{fig:rates-Hdep}
by the solid and dashed lines, respectively.
Clearly the solid line in Fig.~\ref{fig:rates-Hdep},
which represents $\tau_0^{-1}$, explains well
the corresponding experimental curve given in Fig.~\ref{fig:Fdep}.
This means that the strong-collision regime is valid.
The solid line in Fig.~\ref{fig:Tdep} represents the result
for qualitative fitting of
the calculated curve of Eq.~\eqref{eq:T2-B} for zero field to the
experimental temperature dependence
of $T_{2}^{-1}$.
It should be noted that
there appears no definite
dependence of $T_{2}^{-1}$ on the site of the manganese nuclei,
although there exists rather large difference in the $A_{zz}$ component
of the hyperfine tensor as estimated in the previous section (see
Eqs.~(\ref{eq:Atensor})). This is well understood if we adopt
Eq.~\eqref{eq:T2-B}.
The results for the similar fittings for the field
dependence of $T_{2}^{-1}$ are shown in Fig.~\ref{fig:Fdep} by
the solid and dotted lines. The solid line in Fig.~\ref{fig:compFdep}
represents the fitted curve of Eq.~\eqref{eq:T2-B}. In both cases of the
temperature
and field dependence of $T_{2}^{-1}$,
the agreements are  satisfactory.
Thus it is concluded that the transverse
relaxation is essentially determined
by the phase disturbance associated with the average number of the appearance
of the first excited state per one second (strong collision regime).

\subsection{Longitudinal relaxation rate\label{subsec:Analsys T1}}

Now let us turn to the longitudinal relaxation rate.
if we pay attention to the experimental fact that
$T_2^{-1}\gg T_1^{-1}$ together with
the validity of Eq.~\eqref{eq:T2-B},
it is reasonable to assume that the
longitudinal relaxation time $T_1$ is much longer than the
characteristic times $\tau_0$ and $\tau_1$.
Then, the longitudinal relaxation rate $T_{1}^{-1}$
should be obtained, following to the conventional perturbation
theory, by the spectral component 
at the $^{55}$Mn Larmor frequency $\omega_N$ of the time
correlation function for the step-wise fluctuating field as given in
Fig.~\ref{fig:fluc}. The time correlation function for such a  fluctuating
field is
easily calculated to be
\begin{equation}
\<\{h_+(t)h_-(0)\}\>=h_{\rm eff}^2\exp(-\frac{t}{\tau_c})\label{eq:hfluc}
\end{equation}
with
\begin{equation}
 h_{\rm eff}^2=\frac{\tau_o\tau_1}{(\tau_o+\tau_1)^2}h_{\perp}^2\approx
\frac{\tau_1}{\tau_0}h_{\perp}^2\label{eq:heff}
\end{equation}
and
\begin{equation}
 \frac{1}{\tau_c}=\frac{1}{\tau_0}
     +\frac{1}{\tau_1}\approx\frac{1}{\tau_1},\label{eq:tauc}
\end{equation}
where $\tau_c$  is the correlation time.
The approximation used in Eqs.~(\ref{eq:heff}) and (\ref{eq:tauc}) is valid
under the condition that $\tau_0\gg\tau_1$, which is the relevant case.
Taking the
Fourier transform of Eq.~\eqref{eq:hfluc},
we obtain for $\tau_0\gg\tau_1$,
\begin{equation}
\frac{1}{T_1}=\frac{\tau_1}{\tau_0}(\gamma_N h_{\perp})^2
\frac{2\tau_1}{1+(\omega_N\tau_1)^2}\label{eq:T1gen}.
\end{equation}
This equation yields for $\omega_N\tau_1\ll1$
\begin{subequations}\label{eq:T1}
\begin{equation}
\frac{1}{T_1}=\frac{2(\gamma_Nh_\perp\tau_1)^2}{\tau_0}
            \sim \frac{\tau_1^2}{\tau_0},\label{eq:T1-A}
\end{equation}
and for $\omega_N\tau_1\gg1$
\begin{equation}
\frac{1}{T_1}=\frac{2(\gamma_Nh_\perp)^2}{\tau_0\omega_N2}\sim
\frac{1}{\tau_0\omega_N^2}.\label{eq:T1-B}
\end{equation}
\end{subequations}
Here the resonance frequency is given by
$\omega_N=\gamma_N(H_{{\rm int}}\pm H_0)$,
where the signs  $\pm$
correspond to the upper and lower branches for Mn$\uf$ ion and
vice versa for 
Mn$\ut$ ion.
As in the case of $T_{2}^{-1}$, the
temperature dependence of
$T_{1}^{-1}$ is almost determined by $\tau_0$, while the field
dependence depends not only on $\tau_0$ but also on $\tau_1^2$
and $\omega_N^2$. The field dependence of $T_{1}^{-1}$ is determined by
$\tau_1^2\tau_0^{-1}$ for
the low-frequency limit $\omega_Nh_z\tau_1\ll1$, and
by $\tau_0^{-1}\omega_N^{-2}$
for the high-frequency limit $\omega_Nh_z\tau_1\gg1$.
The dotted line in Fig.~\ref{fig:rates-Hdep} represents
the field dependence of $\tau_0^{-1}\omega_N^{-2}$
calculated for $T=1.45$\,K.
 It is found that the experimental
field dependence obtained for $T=1.45$\,K
fits well the dotted line, but not the dashed line,
thus suggesting the
validity of the equation for the high-frequency limit instead of the other.
The dot-dashed and dotted lines in Fig.~\ref{fig:Tdep} represent the results of
the best fitting of the curve $\tau_0^{-1}$ for zero field to
the experimental results for temperature dependence for the  L1-line
(Mn$\uf$ ion)
and the L2-line (Mn$\ut$ ion).
The agreement is reasonable  down to around 0.7\,K.
The results for the similar fittings for the field
dependence of $T_{1}^{-1}$ are shown in Fig.~\ref{fig:Fdep} by
the solid and dotted lines.
The agreement is also satisfactory.
Thus it turns out  that
the longitudinal relaxation is
governed by
perturbing effect of the fluctuating field $h_{\perp}(t)$ (weak collision
model),
and then
the high-frequency limit for the
$\omega_N$
component of the correlation function of $h_{\perp}(t)$ holds.
It should be  noted that the slight difference in the field dependence between
$T_{2}^{-1}$ and $T_{1}^{-1}$ results from the presence
of the factor $\omega_N^{-2}$ in the latter,
which is approximated as ($\gamma_NH_{{\rm int}})^{-2}(1-2H_0/H_{{\rm
int}}$) for $H_{{\rm int}} \gg H_0$.


\section{Discussions\label{sec:discussions}}

First we examine the above treatment numerically. The use of $T_2$
yields directly the value of $\tau_0$. Then we obtain the constant for the spin-phonon interaction,
$C\approx 5\times 10^3$\,s$^{-1}$K$^{-3}$, which lies reasonably in
the range of the  value of 10$^3 \sim10^5$\,s$^{-1}$K$^{-3}$predicted in Ref.~\onlinecite{VBSP}.
Using this value, we obtain $\tau_1\approx1.1\times10^{-7}$\,s.
Here if we tentatively assume that the deviation of the cluster
spin of $\delta S_z$=1 during $\tau_1$ is shared by the 12 manganese spins,
the average deviation of each spin
is taken to be $\delta S_z =1/4$.
Then, for instance,
the use of $A_{zz}/\hbar= 154$\,MHz for Mn$\uf$ ion yields
$\gamma_Nh_z\tau_1\approx$25,
thus the condition for the strong collision regime for $T_{2}^{-1}$
relaxation process being satisfied.
The reason why there appears above about 2K the slight difference
in the value of $T_{2}^{-1}$
between Mn$\uf$ and Mn$\ut$ ions might be due to the cross-over from
the strong-collision regime to the weak collision regime.
In fact, in the latter, the difference in the coupling constant term,
that is, $\gamma_Nh_{z}$ or $A_{zz}/\hbar$ should reflects in the value of
$T_2$.
Nevertheless, as far as we are confined ourselves within the present
simplified treatment, it is not realized because $\tau_1$ is almost
temperature-independent.

As for $T_1$, for instance,
using the evaluated value of $A_{xz}/\hbar =24.7$\,MHz (Mn(2)) and 53\,MHz (Mn(3)),
we obtain
$T_2^{-1}/T_1^{-1} \approx 500$  and  $100$,
respectively,
which agrees reasonably with the experimental result
in relative order of magnitude.
However, the following points
remain not understood.
First, as far as the hyperfine interaction for Mn$\uf$ ion is taken to be isotropic
and not to have off-diagonal terms as determined in the present analysis of
the NMR spectra,
it is difficult to understand that the relaxation rate for Mn$\uf$ ion
is of comparable order with that for Mn$\ut$ ion.
In order to understand
the experimental results for $T_1$ for Mn$\uf$ ion, 
the assumption of
the presence of any off-diagonal term
of the hyperfine interaction tensor
will be necessary.
Unless the anisotropic term is not effective, the possible relaxation mechanism
may be inevitably ascribed to the isotropic interaction term
like $A I^{+}$$\delta$$S^{-}(t)$ as in the case of
the three-magnon relaxation process in usual magnetic system.
In this case,
the relevant activation energy which determines the temperature
dependence of $T_1$ should be at least twice of the gap energy so as to
guarantee the energy conservation between the nuclear spin and the
electronic spin system.
However, in the cluster which involves only
twelve electronic spins coupled strongly to each other by  exchange
interactions, it may be quite unreliable.
Anyhow, the origin of the effective coupling constant for $T_1$ for Mn$\uf$
ion is
uncertain at the present.

Next we discuss on  the present concept for obtaining the
relevant equations for $T_2$ and $T_1$. As a starting point we
considered only the two lowest levels,
the ground-state and the first excited state.
Then, if we follow the standard stochastic theory, the
fluctuating local field responsible for the nuclear magnetic
relaxation, which is caused by the thermal excitation to the excited
state from the ground-state, is treated as a perturbation with
respect to the nuclear quantization axis. In the cases of zero field
and where the external field is applied along the c-axis, the nuclear
quantization axis is taken to be along the internal field at the
$^{55}$Mn site, that is, along the $c$-axis. The longitudinal and
transverse relaxation rates are given by the Fourier spectrum of
correlation function of such a fluctuating local field
$h_{\alpha}(t)$ ($\alpha=z$ or $\perp$) at the resonance
frequency $\omega_N$ as follows:
\begin{equation*}
\frac{1}{T_1}=F_{\perp}(\omega_N)   \quad \text{and}  \quad
\frac{1}{T_2}=\frac{1}{2T_1}+F_z(0)
\end{equation*}
with
\begin{equation}
F(\omega_N)=\frac{\gamma_N^2}{2}\int_{-\infty}^{\infty}
\<\{\delta h_{\alpha}(t)\delta h_{\alpha}(t)\}\>\exp(-i\omega_Nt)dt.
\end{equation}
As shown in Sec.~\ref{sec:An}, the correlation function for the pulse-like
field given in Fig.~\ref{fig:fluc},
is calculated to have the exponential-typed
form given by Eq.~\eqref{eq:hfluc}. The longitudinal relaxation rate has
already
been given by Eq.~\eqref{eq:T1gen}. On the other hand, we have found
experimentally
that $T_2^{-1}\gg T_1^{-1}$. So $T_2^{-1}$ should be
ascribed to the zero-frequency term $F(0)$,  which is given by
\begin{equation}
\frac{1}{T_2}=F(0)=\frac{\tau_1^2}{\tau_0}(\gamma_Nh_z)^2.
\end{equation}
According to this equation, the temperature dependence of $T_{2}^{-1}$
results only from  $\tau_0$ as far as $\tau_0\gg\tau_1$ because $\tau_1$ is
taken to be almost independent on the temperature.
However, the value of $T_{2}^{-1}$  should have a large site-dependence
through the
coupling constant term $(\gamma_Nh_z)^2$, which is proportional to
the hyperfine interaction term of $A_{zz}^2$. Furthermore, as already
shown in Fig.~\ref{fig:rates-Hdep}, the field dependence of
$\tau_1^2\tau_0^{-1}$  differs from that of only $\tau_0^{-1}$.
Such features contradict with our experimental results. Instead, as
we have already mentioned, the equation like
$T_2^{-1}=\tau_0^{-1}$  obtained as a strong collision regime on
the basis of more basic treatment explains satisfactorily our
experimental results for $T_2$.

Finally, with respect to the longitudinal relaxation, 
we compare our
present formalism with the equation adopted by Furukawa \etal. in
Ref.~\onlinecite{FWKBG}.
They have taken into account all of the 21 energy
levels ($m=-10\sim+10$) of $S=10$ as
the candidates for the fluctuating field
$\delta h_{\perp}$
responsible for the nuclear magnetic relaxation.
Then, it has been assumed
that the correlation functions associated with each of these energy
levels are the exponential-typed one  with the correlation time equal
to the corresponding  average life time, which is  determined by the
spin-phonon interaction as discussed  in Ref.~\onlinecite{VBSP}.
The contributions from each level have been summed up with
the statistical weight.
However, in considering  the low temperature results,
only the term
related to the lowest ground-state with $m=-10$
with the predominant statistical weight
was remained as an effective one,
thus yielding the expression like
$T_1^{-1}\propto A_{\pm}^2/\tau_0\omega_N^2$ in the
high-frequency limit for $\tau_0\omega_N\gg1$.
As is seen, our equation ~\eqref{eq:T1-B} has the
same form as this equation.
However, the origins of each term are different.
First, in Eq.~\eqref{eq:T1-B}, the term $\tau_0$
results from the effective amplitude of the fluctuating field
$(\tau_1/\tau_0)(\gamma_Nh_{\perp})^2$  in the correlation
function.
Secondly the criterion for the high-frequency limit,
which brings the term $\omega_N$ in the equation, is taken with respect to
$\tau_1\omega_N$ instead of $\tau_0\omega_N$.
Thirdly, as for the coupling constant, Eq.~\eqref{eq:T1-B} involves
the transverse fluctuating field $h_{\perp}$,
that is, the off-diagonal term $A_{xz}$,
since it is assumed that the
the anisotropic perturbing interaction like
$I^{+}\delta S_z(t)$ is responsible for the nuclear spin lattice relaxation.
While, in Ref.~\onlinecite{FWKBG},
the coupling constants for Mn(1) , Mn(2), and Mn(3) 
have been taken to be proportional to
the square of
the internal static field in zero field, \ie, square of $A_{zz}$ in our
notation.
Then,
in view of the large difference of the ratio
of the coupling constants of $T_1^{-1}$ for the three ions
between experimental values
and the above estimation, the presence of the coupling constants for
non-zero mode have been suggested.
Unfortunately the $^{55}$Mn
NMR signal is observable only at low temperatures. So it is
difficult to find experimentally the maximum point of the relaxation
rate which appears in the BPP-typed equation under the condition
$\tau_1\omega_N=1$.
It may be possible for other nuclei with much lower resonance frequency.
However, such a condition is realized
at rather high temperatures. In this case,
the present two-level model may fail.
Further extension of the present treatment will be
necessary by taking into account the higher excited levels.

\section{Conclusion}
The nuclear magnetic relaxation times $T_2$ and $T_1$ of
 $^{55}$Mn for Mn$\uf$ and Mn$\ut$ ions
in Mn$_{12}$Ac have been measured using the oriented powder sample
at low temperatures below 2.5\,K down to 200\,mK in the fields 
up to 9\,T applied along $c$-axis.
The relaxation rates $T_{2}^{-1}$ and $T_{1}^{-1}$ 
in zero field exhibited remarkable
decreases with decreasing temperature 
with the
relative relation like $T_2^{-1}/T_1^{-1}=200$ 
at the temperatures above about 1.5\,K. 
At the lower temperatures the difference was more pronounced.
The field dependence of $T_2^{-1}$ and $T_1^{-1}$ showed 
decrease with increasing field for the cluster whose
magnetic moment is parallel to the $c$-axis.
On the contrary, those for the cluster whose 
magnetic moment is antiparallel to the $c$-axis increase
with increasing field.
The analysis for the experimental results
was made on the concept that the nuclear magnetic relaxation is
caused by thermal excitations of the cluster spin. 
We simplified
the problem by considering only the excitation 
from the ground-state to the first excited state 
within each well of the double-well potential 
with the average life times determined by the spin-phonon interaction.
On basis of the nonlinear theory,
we obtained general expression for the transverse relaxation rate $T_2^{-1}$ for such a pulse-like fluctuating field.
It turned
out that $T_{2}^{-1}$ is well understood in terms of 
the equation $T_2^{-1}=\tau_0^{-1}$, which corresponds to 
the strong collision regime.
On the other hand, the experimental results
for $T_1^{-1}$ was interpreted well by the standard
perturbation formalism.
On the other hand, the results for $T_1$ have
been well understood, on the basis of the standard
perturbation method, by the equation for the the high-frequency limit 
like $T_1^{-1}\sim 1/\tau_0 \omega_N^2$, where 
$\omega_N$ is the $^{55}$Mn Larmor frequency.  
The quantitative comparison between the experiment 
and the theoretical calculation
was made by using the $^{55}$Mn hyperfine interaction tensors,
which resulted in reasonable agreement.


\section*{Acknowledgments}
The authors wish to thank Professor T. Kohmoto for valuable discussions.
Discussions with Professor F. Borsa 
and Dr. A. Lascialfari are greatfully appreciated.
Thanks are also due to Dr. Y. Furukawa for giving us
useful informations.
This work is supported by the Grant-in-Aid for Scientific Research from
Ministry of Education, Culture, Sports, and Technology.


\begin{thebibliography}{99}

\bibitem{L}
T.~Lis,  Acta.
  Cryst. \textbf{B36}, 2042
  (1980).

\bibitem{SGCN}
R.~Sessoli, 
  D.~Gatteschi, 
  A.~Canesschi, 
and  M.~A. Novak, 
  Nature \textbf{365},
  (1993).

\bibitem{VBSP}
J.~Villain, 
  F.~Hartoman-Boutron, 
  R.~Sessoli,  and
  A.~Pettori, 
  Europhys. Lett. \textbf{27},
  159 (1994).

\bibitem{TLBGSB}
L.~Thomas, 
  F.~L. Lionti, 
  R.~Ballou, 
  D.~Gatteschi, 
  R.~Sessoli,  and
  B.~Barbara, 
  Nature \textbf{383},
  145 (1996).

\bibitem{FSTZ}
J.~R. Friedman, 
  M.~P. Sarachik, 
  J.~Tejada,  and
  R.~Ziolo, 
  Phys. Rev. Lett. \textbf{76},
  3830 (1996).

\bibitem{PRHV}
P.~Politi, 
  A.~Rettori, 
  F.~Hartmann-Boutron, 
and  J.~Villain, 
  Phys. Rev. Lett. \textbf{75},
  537 (1995).
  
\bibitem{SOPSG}
C.~Sangregorio, 
  T.~Ohm, 
  C.~Paulsen, 
  R.~Sessoli,  and
  D.~Gatteschi, 
  Phys. Rev. Lett. \textbf{78},
  4645 (1997).

\bibitem{WSG}
W.~Wernsdorfer, 
  R.~Sessoli,  and
  D.~Gatteschi, 
  Science \textbf{184},
  133 (1999).

\bibitem{CWMBB}
I.~Chiorescu, 
  W.~Wernsdorfer, 
  A.~M\"{u}ller, 
  H.~Bogge,  and
  B.~Barbara, 
  Phys. Rev. Lett. \textbf{84},
  3454 (2000).


\bibitem{GB}
L.~Gunther and
  B.~Barbara, 
  \emph{Quantum Tunneling of Magnetization}
  (Kluwer, Dordrecht, 1995).

\bibitem{CT}
E.~M. Chudnovsky
  and J.~Tejada, 
  \emph{Macroscopic Quantum Tunneling of Magnetic Moment}
  (Cambridge University Press, Cambridge,  1997).

\bibitem{HPV}
F.~Hartman-Boutron, 
  P.~Politi,  and
  J.~Villain, 
  J. Mod. Phys. \textbf{B21},
  2577 (1996).

\bibitem{TPS}
L.~Tupitsyn and
  N.~V.~P. and P. C.~E.~Stamp, 
  J. Mod. Phys. \textbf{B11},
  2901 (1997).
  
\bibitem{TB}
I.~Tupitsyn and B.~Barbara,
  \emph{Quantum tunneling of magnetization in molecular complexes with large spins. Effect of the environment.}
, \texttt{cond-mat/0002180}.

\bibitem{GKKFOATA}
T.~Goto, 
  T.~Kubo, 
  T.~Koshiba, 
  Y.~Fujii, 
  A.~Oyamada, 
  J.~Arai, 
  K.~Takeda,  and
  K.~Awaga, 
  Physica B \textbf{284},
  1277 (2000).

\bibitem{KGKTA}
T.~Kubo, 
  T.~Goto, 
  T.~Koshiba, 
  K.~Takeda,  and
  K.~Awaga, 
  Phys. Rev. B \textbf{65},
  224425 (2002).

\bibitem{LGBSJC}
A.~Lascialfari, 
  D.~Gatteschi, 
  F.~Borsa, 
  A.~Shastri, 
  Z.~H. Jang,  and
  P.~Carretta, 
  Phys. Rev. B \textbf{57},
  514 (1998-I).

\bibitem{LJBCG}
A.~Lascialfari, 
  Z.~H. Jang, 
  F.~Borsa, 
  P.~Carretta,  and
  D.~Gatteschi, 
  Phys. Rev. Lett. \textbf{81},
  3773 (1998).

\bibitem{FWKBG}
Y.~Furukawa, 
  K.~Watanabe, 
  K.~Kumagai, 
  F.~Borsa,  and
  D.~Gatteschi, 
  Phys. Rev. B \textbf{64},
  104401 (2001).

\bibitem{KGKA}
T.~Koshiba, 
  T.~Goto, 
  T.~Kubo,  and
  K.~Awaga, 
  Prog. Theor. Phys. Suppl. \textbf{145},
  2002 (2002).

\bibitem{STSWVFGCH}
R.~Sessoli, 
  H.~L. Tsai, 
  A.~R. Schake, 
  S.~Wang, 
  J.~B. Vincent, 
  K.~Folting, 
  D.~Gatteschi, 
  G.~Christou,  and
  D.~N. Hendricson, 
  J. Am. Chem. Soc. \textbf{115},
  1804 (1993).

\bibitem{CGSBBG}
A.~Caneschi, 
  D.~Gatteschi, 
  R.~Sessoli, 
  A.~L. Barra, 
  L.-D. Brunel, 
and  M.~Guillot, 
  J. Am. Chem. Soc. \textbf{113},
  5873 (1991).

\bibitem{RJSGV}
N.~Regnault, 
  T.~Jolic\oe ur, 
  R.~Sessoli, 
  D.~Gatteschi, 
and 
  M.~Verdaguer, 
  \emph{Exchange couplings in the magnetic molecular cluster}
  Mn$_{12}${A}c, \texttt{cond-mat/0203480}.

\bibitem{BGS}
A.~L. Barra, 
  D.~Gatteschi, 
and  R.~Sessoli, 
  Phys. Rev. B \textbf{56}, 
  8192 (1997).

\bibitem{MHCAGIC}
I.~Mirebeau, 
  M.~Hennion, 
  H.~Casalta, 
  H.~Andres, 
  H.~U. Gudel, 
  A.~V. Irodova, 
and  A.~Caneschi, 
  Phys. Rev. Lett. \textbf{83},
  628 (1999).

\bibitem{PS}
T.~Pohjola and
  H.~Schoeller, 
  Phys. Rev.B \textbf{62},
  15026 (2000-II).

\bibitem{FW}
A.~J. Freeman and
  R.~E. Watson,
  \emph{Magnetism IIA},  edited by T. Rado and H. Shull
  (Academic Press, New York, 165).

\bibitem{RBAHA}
A.~Robinson, 
  P.~J. Brown, 
  D.~N. Argyriou, 
  D.~N. Hendrikson, 
and  M.~J. Aubin, 
  J. Phys. B \textbf{12},
  2805 (2000).

\bibitem{AT}
P.~W. Andrew and
  D.~P. Tunstall, 
  Proc. Phys. Soc. \textbf{78},
  1 (1961).

\bibitem{VBSR}
J.~Villan, 
  F.~H. Boutron, 
  R.~Sessoli,  and
  A.~Rettori, 
  Euro. Phys. Lett. \textbf{27},
  159 (1994).

\bibitem{KA}
J.~R. Klander and
  P.~W. Anderson, 
  Phys. Rev. \textbf{125},
  912 (1962).

\bibitem{KGMFFKM}
T.~Kohmoto, 
  T.~Goto, 
  S.~Maegawa, 
  N.~Fujiwara, 
  Y.~Fukuda, 
  M.~Kunitomo,  and
  M.~Mekata, 
  Phys. Rev. B \textbf{49},
  6028 (1994).
\end{thebibliography}


\end{document}